\documentclass[10pt]{iopart}
\usepackage{amsthm,amssymb,euscript,amscd}
\usepackage{graphicx,xspace,colortbl}
\newif\ifpdflatex\pdflatextrue
\makeatletter\@ifundefined{pdfoutput}{\pdflatexfalse}\makeatother
\newif\ifscreenformat\screenformattrue
\def\myincludegraphics[#1]#2#3{%
    \ifpdflatex \includegraphics[#1]{#2}
    \else       \includegraphics[#1]{#3}
    \fi}
\ifpdflatex 
    \usepackage[pdftex,backref,colorlinks]{hyperref}
    \hypersetup{%
      pdfauthor  = {Sonia G Schirmer},
      pdftitle   = {Necessary and sufficient criteria for simulataneous controllability},
      pdfsubject = {Quantum control},
      pdfkeywords= {controllability, optimal control, quantum systems, quantum dots}}
\else  
    \usepackage[dvips]{hyperref}
\fi

\def\op#1{#1}        
\def\op#1{\hat{#1}}  
\def\vec#1{\mathbf{#1}}
\def\ket#1{| #1 \rangle}
\def\bra#1{\langle #1 |}

\def\diag{\mbox{\rm diag}}

\def\Tr{\mathop{\rm Tr}}
\def\dim{\mathop{\rm dim}}

\def\eps{\epsilon}
\def\sx{\op{\sigma}_x}
\def\sy{\op{\sigma}_y}
\def\sz{\op{\sigma}_z}

\def\H{{\cal H}}
\def\L{{\cal L}}

\def\u{{\rm u}}
\def\su{{\rm su}}
\def\so{{\rm so}}
\def\sp{{\rm sp}}
\newtheorem{theorem}{Theorem}

\begin{document}
\title[Controllability of multi-partite quantum systems]{Controllability of multi-partite
quantum systems and selective excitation of quantum dots}
\author{S.~G.\ Schirmer}
\address{Dept of Applied Maths and Theoretical Physics,
         University of Cambridge, Wilberforce Road, Cambridge, CB3 0WA, United Kingdom}
\ead{\mailto{sgs29@cam.ac.uk}}
\author{I.~C.~H.\ Pullen}
\address{Dept of Applied Maths and Computing, The Open University, Walton Hall, 
         Milton Keynes, MK7 6AA, United Kingdom}
\author{A.~I.\ Solomon}
\address{Dept of Physics and Astronomy, The Open University, Walton Hall, 
         Milton Keynes, MK7 6AA, United Kingdom}
\date{\today}
\begin{abstract}
We consider the degrees of controllability of multi-partite quantum systems, as well as
necessary and sufficient criteria for each case.  The results are applied to the problem
of simultaneous control of an ensemble of quantum dots with a single laser pulse.  Finally, 
we apply optimal control techniques to demonstrate selective excitation of individual dots
for a simultaneously controllable ensemble of quantum dots.
\end{abstract}

\section{Introduction}

Control of quantum processes is essential to realize the vast potential of quantum 
technology ranging from applications in quantum information processing~\cite{00Nielsen} 
to atomic and molecular physics and chemistry~\cite{00Rice}.  Although robust and 
efficient control of quantum phenomena remains a challenge, especially in practice, 
recent advances in theory and technology have made control of systems at the quantum 
level increasingly feasible~\cite{SCI288p0824}.  

To achieve the best possible control outcomes, it is crucial to know the degree to 
which a quantum system is controllable, to understand fundamental limits on control,
and to find optimal ways to implement control given certain constraints.  Although 
significant progress has been made recently with regard to refining the notions of 
and criteria for controllability of quantum systems~\cite{JMP24p2608,PRA51p0960,
PRA62n022108,CP267p001,IEEE39CDC1086,IEEE39CDC2803,IEEE39CDC3003,PRA63n063410,
JPA34p1679,qph0106128,JPA35p4125,JMP43p2051,qph0302121}, understanding constraints%
~\cite{PRA58p2684,PRA63n025403,IJMPB16p2107}, and developing techniques for optimal 
control field design~\cite{JCP88p6870,JCP92p364,JCP93p1670,PRL68p1500,JPC97p2320,
JCP100p1094,PRA52pR891,CPL259p488,JCP109p0385,JCP110p9825,PRA61n012101,PRE70n016704}, 
many interesting questions remain.  

Among these are the controllability and optimal control of multi-partite systems, i.e., 
systems comprised of $L$ distinct quantum elements such as quantum dots or molecules.  
The obvious questions to ask in this case include (a) whether each of the components 
is controllable individually, and (b) whether the composite system is controllable as
a whole.  In practice, however, one often faces more subtle questions.  For instance, 
given an ensemble of quantum dots~\cite{99Bimberg}, with negligible interdot coupling
but clustered too close together to allow (spatial) addressing of individual dots, 
when is it possible to selectively excite individual dots in the ensemble?  What are
appropriate notions of controllability?  How can we achieve selective excitation in 
practice?  

To address these questions, we first review existing notions of controllability for 
(finite-dimensional) quantum systems (Sec.~\ref{sec:notions}).  We then proceed to 
consider the application of these notions to multi-partite systems, introduce the
notion of simultaneous controllability  (Sec.~\ref{sec:notions-multi}), and discuss 
necessary and sufficient criteria for the latter notion (Sec.~\ref{sec:criteria}).  
Finally, the results will be applied to the problem of simultaneous controllability
of ensembles of quantum dots (Sec.~\ref{sec:dots_controllability}) and selective 
excitation of individual dots using optimal control (Sec.~\ref{sec:dots_optimal}).

\section{Notions of Controllability for Quantum Systems}
\label{sec:notions}

We will restrict ourselves here to discussing basic notions of controllability for 
finite-dimensional, Hamiltonian quantum systems subject to open-loop coherent control, 
i.e., a system whose evolution is governed by the quantum Liouville equation
\begin{equation} \label{eq:LE}
  \rmi\hbar \frac{d}{d t} \op{\rho}(t) 
   = \op{H}[\vec{f}(t)]\op{\rho}(t) - \op{\rho}(t)\op{H}[\vec{f}(t)],
\end{equation}
where $\op{\rho}(t)$ is the density operator representing the state of the system, 
and $\op{H}[\vec{f}(t)]$ is the total Hamiltonian, which depends on various control 
fields $\vec{f}=(f_1,\ldots,f_M)$, e.g.,
\begin{equation} \label{eq:Ht}
  \op{H}[\vec{f}(t)] = \op{H}_0 + \sum_{m=1}^M f_m(t) \op{H}_m.
\end{equation}
The strongest possible requirement in terms of controllability, sometimes referred 
to as \emph{complete controllability}~\cite{PRA63n063410}, is the ability to create
any dynamical evolution, which for a Hamiltonian system is equivalent to the ability 
to dynamically generate any unitary operator in $U(N)$, where $N$ is the dimension 
of the system's Hilbert space, by applying a suitable control field $\vec{f}(t)$.  

Another common notion is \emph{mixed-state controllability}, which requires that the
system can be driven from any given (pure or mixed) initial state $\op{\rho}_0$, to 
any kinematically equivalent state $\op{\rho}_1=\op{U}\op{\rho}_0\op{U}^T$, $\op{U}$ 
being a unitary operator in $U(N)$, by applying a suitable control field $\vec{f}(t)$. 
Mixed-state controllability also implies observable controllability~\cite{JPA35p4125}.

If the initial state of the system is known to be pure, one may instead consider the
weaker requirement of \emph{pure-state controllability}---also referred to as wave-%
function controllability~\cite{CP267p001}---which simply requires that the system can 
be driven from any pure initial state $\op{\rho}_0=\ket{\Psi}\bra{\Psi}$ to any other 
pure state by applying a suitable control field $\vec{f}(t)$.  

Considerable work has been done to find necessary and sufficient conditions for each
of these concepts of controllability, especially in terms of the dynamical Lie algebra 
$\L$ generated by the skew-Hermitian matrices $\rmi H_m$, $m=0,\ldots M$ (see e.g.~%
\cite{JMP24p2608,PRA51p0960,qph0106128,JPA35p4125,JMP43p2051}).  The key results can 
be summarized in the following theorem:
\begin{theorem}
{\bf (Degrees of Controllability)}\label{thm:degCtrl} 
The system defined by Eqs~(\ref{eq:LE}), (\ref{eq:Ht}) is 
\begin{itemize}
\item completely controllable if and only if $\L \equiv \u(N)$, the Lie algebra of 
      skew-Hermitian $N \times N$ matrices;
\item mixed-state controllable if and only if $\L\equiv \u(N)$ or $\L\equiv \su(N)$, 
      the Lie algebra of traceless skew-Hermitian $N \times N$ matrices;
\item pure-state controllable if and only if $\L$ is $\u(N)$, $\su(N)$, or if $N$ is 
      even, the symplectic Lie algebra $\sp(N/2)$ or $\sp(N/2)\oplus \u(1)$.
\end{itemize}
\end{theorem}
Noting that $\dim_R \u(N)=N^2$, $\dim_R \su(N)=N^2-1$ and $\dim_R \sp(N/2)=N(N+1)/2$, 
these Lie-algebraic criteria allow us to easily verify the degree of controllability 
of a system by simply computing the dimension of the dynamical Lie algebra.  

Mixed-state controllability appears to be the most important notion of controllabilty
as it implies pure-state controllability for Hamiltonian systems, and unlike pure-state
controllability, can be extended to open systems and systems subject feedback control%
---although the definition needs to be modified slightly in the latter cases.  Hence,
we shall refer to mixed-state controllable systems simply as controllable in the
following.  For a discussion of notions of controllability for quantum systems subject
to feedback control we refer the reader to~\cite{PRA62n022108}.  For open systems whose
evolution is governed by a dynamical semi-group we draw attention to the related concept
of the set of states that are reachable from a given initial state at a certain target 
time by applying suitable control fields~\cite{JMP24p2608,JPA35p8551}.

\section{Notions of Controllability of Multi-partite Systems}
\label{sec:notions-multi}

The degrees of controllability defined above obviously apply to any quantum system,
including composite or multi-partite systems.  However, in the latter case it is 
useful to introduce additional notions of controllability.  

For instance, consider a system of $L$ particles or quantum units such as different 
molecules or quantum dots.  If the Hilbert space of particle $\ell$ is $\H_\ell$ then
the Hilbert space of the composite system is usually the tensor product space $\H = 
\H_1 \otimes \ldots \otimes \H_L$, and if $\dim \H_\ell=N_\ell$ then the dimension of
the tensor product space is $N = N_1 \cdots N_L$.  Hence, by Theorem~\ref{thm:degCtrl} 
the composite system is controllable if and only if the dynamical Lie algebra generated
by the system and control Hamiltonians $\op{H}_0$ and $\op{H}_m$, $m = 1,\ldots,M$, is 
$\u(N)$ or $\su(N)$.  However, the tensor product space may be huge and not always the
the most appropriate Hilbert space for the system.

If the particles do not interact with each other, for example, then we can reduce the
Hilbert space of the composite system to the direct sum of the Hilbert spaces $\H_\ell$
of the non-interacting parts rather than the tensor product.  This situation arises in 
practice in quantum chemistry, in particular laser control of chemical reactions, where 
we may want to control several different types of molecules in a dilute solution, all 
of which interact simultaneously with an external control field such as a laser pulse,
but for which intermolecular interactions are negligible~\cite{JPCB106p8125}.  The same
problems are encountered for ensembles of quantum dots close enough together to prevent
selective addressing of a single dot with a laser, but with negligible interdot coupling 
(e.g., due to different dot sizes etc.)  We will refer to this case of a quantum system 
consisting of multiple non-interacting quantum units as \emph{decomposable} or 
\emph{separable}.  Clearly, such a system is not controllable or even pure-state 
controllable in the stong sense defined in Sec.~\ref{sec:notions}.

Appropriate notions of controllability in this case are the individual controllability
of the components (separate quantum units) and their simultaneous controllability.  The
latter is a stronger notion than the former as individual controllability of each unit
does not imply that we can \emph{simultaneously} control all units.  For example, even
if we could completely control each dot in a cluster of quantum dots individually, this
does not necessarily mean that we can simultaneously control all the dots in the cluster
with the same control pulse or pulse sequence.  This naturally prompts the question as 
to when the components of a separable quantum system are simultaneously controllable.

\section{Criteria for Simultaneous Controllability}
\label{sec:criteria}

The Hilbert space of a quantum system that consists of $L$ non-interacting subsystems 
can be represented as the direct sum $\H=\H_1\oplus \ldots \oplus\H_L$ of the Hilbert
spaces $\H_\ell$ of the independent subsystems, $\ell=1,\ldots,L$.  Thus, with respect
to a suitable basis, the state $\op{\rho}(t)$ and the Hamiltonians $\op{H}_m$, $0 \le 
m\le M$, have a block-diagonal structure
\begin{equation} \label{eq:Hm}
  \begin{array}{rcl}  
   \op{\rho}(t) &=& \diag(\op{\rho}_1(t),\ldots,\op{\rho}_L(t))  \\
   \op{H}_m     &=& \diag(\op{H}_{m,1},\ldots,\op{H}_{m,L}), \quad m=0,1,\ldots,M
  \end{array}
\end{equation}
where $\op{\rho}_\ell(t)$ and $\op{H}_{m\ell}$ are $N_\ell \times N_\ell$ matrices and
$N_\ell = \dim\H_\ell$.  Note that we require at least two Hamiltonians, $\op{H}_0$
and $\op{H}_1$, but in some cases it is necessary or at least advantageous to have
more control Hamiltonians.  For example, given an ensemble of optically controlled
quantum dots that is sufficiently large, it may be possible to focus the laser beam 
on a certain region, which would allow us to define a separate control Hamiltonian 
for each independently accessible region.  In other cases, e.g., for an ensemble of 
quantum dots whose internal energy level structure requires control fields with 
different polarizations, multiple control Hamiltonians may be essential.

It is obvious from this structure of the dynamical generators in Eq.~(\ref{eq:Hm}) 
that the dynamical Lie group of the system must be contained in $U(N_1) \times \ldots 
\times U(N_L)$, and that maximal orbits for a generic mixed state~\cite{JPA37p1389} 
under the action of this Lie group are thus homeomorphic to $U(N_1)\times\ldots\times 
U(N_L)/ [U(1) \times \ldots \times U(1)]$, where there are $N$ terms in the denominator.  
The latter is equivalent to $SU(N_1) \times\ldots\times SU(N_L)/[U(1)\times\ldots\times
U(1)]$ where we have $N-L$ terms in the denominator.  These considerations lead to the 
following criterion (see~\ref{app:proof}):

\begin{theorem}{\bf (Simultaneous Controllability)} \label{thm:simCtrl}
The independent components of a decomposable system with Hamiltonian (\ref{eq:Ht}) 
of the form (\ref{eq:Hm}) are \emph{simultaneously mixed-state controllable} if and 
only if the dimension of the dynamical Lie algebra $\L$ is
\[
   \dim \L = r + \sum_{\ell=1}^L (N_\ell^2-1),
\]
where $r$ is the rank of matrix $A$ of Eq.~(\ref{eq:A}).
\end{theorem}

This result is a generalization of the sufficient criteria in~\cite{JPA37p0273} for 
arbitrary $M$.  Specifically, note that if we have only two Hamiltonians $\op{H}_0$ and
$\op{H}_1$ of the form~(\ref{eq:Hm}), i.e., $M=1$, then we have $r=0$ if the diagonal
generators $\tilde{D}_0$ and $\tilde{D}_1$ defined in \ref{app:proof} are both 0, $r=1$ 
if they are linearly dependent, and $r=2$ if they are linearly independent, i.e., we 
recover the sufficient criteria of Theorem 2 in~\cite{JPA37p0273}.  Furthermore, our 
derivation in \ref{app:proof} also shows that the Lie algebra dimension condition in 
Theorem~\ref{thm:simCtrl} is both necessary and sufficient, at least for generic 
mixed-state controllability.  

Finally, a similar argument shows that a necessary and sufficient condition for the 
weaker notion of \emph{simultaneous pure-state controllability}---which is sufficient
if each subsystem is known to be in a pure quantum state initially and evolves unitarily%
---is that the Lie algebra $\tilde{\L}$ (see \ref{app:proof}) be a direct sum of $L$ terms, 
where the $\ell$th terms is either $\su(N_\ell)$, or if $N_\ell$ is even, $\sp(N_\ell/2)$.  

\section{Controllability of Ensembles of Quantum Dots}
\label{sec:dots_controllability}

Consider again the example of an ensemble of $L$ quantum dots, each of which is itself
an $N_\ell$-dimensional quantum system.  According to Theorem~\ref{thm:degCtrl} each of
the dots is individually controllable if its associated dynamical Lie algebra $\L_\ell$ 
has (at least) dimension $N_\ell^2-1$, and the entire ensemble is controllable as an $N
=N_1 \times \ldots \times N_L$ dimensional composite system if the Lie algebra $\L$ of 
the entire system has (at least) dimension $N^2-1$.  If interdot coupling is negligible 
then the system is decomposable, and hence not controllable as a composite system, but 
its components are simultaneously controllable if the Lie algebra of the system satisfies 
Theorem~\ref{thm:simCtrl}.  

If each dot can be represented as a two-level system with Hamiltonian
\begin{equation}
  \op{H}^{(\ell)}[f(t)] = \eps_\ell \sz + f(t) d_\ell \sx,
\end{equation}
where $\sx$ and $\sz$ are the Pauli matrices, then we can use Theorem~\ref{thm:simCtrl} 
to give explicit conditions for simultaneous controllability.  Clearly, the $\ell$th 
dot is controllable individually exactly if $\eps_\ell \neq 0$ and $d_\ell \neq 0$, as 
$[\sx,\sz]=-2\sy$.  However, this is not sufficient for simultaneous controllability.  
If two or more dots in the ensemble have exactly the same system parameters, for example, 
then the Lie subalgebra they generate is a higher-dimensional representation of $\su(2)$ 
and the dots are not simultaneously controllable.  In this simple case one can show 
(see~\ref{app2}) that the ensemble of dots is simultaneously controllable exactly if 
$(\eps_\ell,d_\ell)\neq (\pm\eps_{\ell'},\pm d_{\ell'})$ unless $\ell=\ell'$.  

\begin{figure}
\center
\myincludegraphics[width=0.5\textwidth]{figures/pdf/Diagram1.pdf}{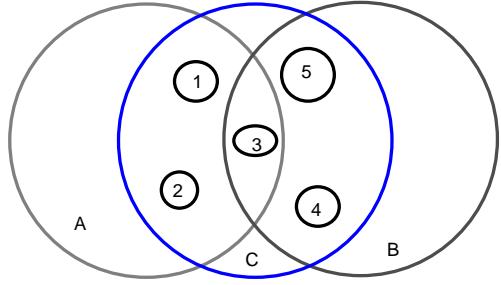}
\caption{Five-dot ensemble: If dots 1 and 4 are identical the ensemble is not 
simultaneously controllable if the laser is centered in position $C$.  However,
if we can create two regions $A$ and $B$, which can be separately addressed by 
the laser, then the ensemble becomes simultaneously controllable.}\label{fig1}
\end{figure}

However, even if two or more dots in the ensembles are identical, we can often recover 
simultaneous controllability if we can divide the ensemble into (possibly overlapping) 
regions that can be selectively addressed.  For example, consider the five-dot ensemble 
shown in Fig.~\ref{fig1}, where dots 1 and 4 are identical but all the other dots are 
distinct, i.e., $(\eps_1,d_1) = (\eps_4,d_4)$, but $(\eps_\ell,d_\ell) \neq (\pm\eps_{\ell'}, 
\pm d_{\ell'})$ for $\ell \neq \ell'$ and $\ell,\ell' \in \{1,2,3,5\}$.  As dots 1 and 4 
have the same characteristics, $\op{H}^{(1)}=\op{H}^{(4)}$, the Lie algebra generated by 
\begin{eqnarray*}
  \op{H}_0 &=& \eps_1\sz \oplus \eps_2\sz \oplus \eps_3\sz \oplus \eps_1\sz \oplus \eps_5\sz\\ 
  \op{H}_C &=& d_1\sx \oplus d_2\sx \oplus d_3\sx \oplus d_1\sx \oplus d_5\sx 
\end{eqnarray*}
has dimension 12, and the ensemble is \emph{not} simultaneously controllable.  However, 
if we can adjust the laser to create two regions $A$ and $B$, encompassing dots $\{1,2,
3\}$ and $\{3,4,5\}$, respectively, simultaneous controllability can be recovered as the
Lie algebra generated by $\{\op{H}_0,\op{H}_A,\op{H}_B\}$ with
\begin{eqnarray*}
  \op{H}_0 &=& \eps_1\sz \oplus \eps_2\sz \oplus \eps_3\sz \oplus \eps_1\sz \oplus \eps_5\sz\\ 
  \op{H}_A &=& d_1\sx \oplus d_2\sx \oplus d_3\sx \oplus 0 \oplus 0\\ 
  \op{H}_B &=& 0 \oplus 0 \oplus d_3\sx \oplus d_1\sx \oplus d_5\sx
\end{eqnarray*}
has dimension 15, as required.

\begin{figure}
\center
\myincludegraphics[width=0.5\textwidth]{figures/pdf/Diagram2.pdf}{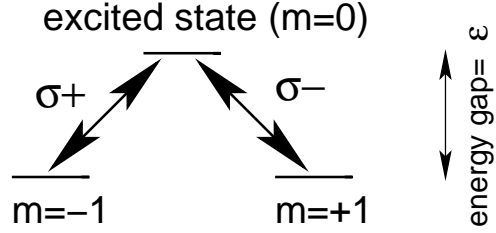}
\caption{Three-state dot in $\Lambda$ configuration.}
\label{fig2}
\end{figure}

We can also apply the controllability results to dots with a more complicated internal
structure.  For instance, consider an ensemble of $L$ dots with a two-fold degenerate 
internal ground state and an excited state in a $\Lambda$ configuration as shown in 
Fig.~\ref{fig2}.  Let $\eps_\ell$ denote the energy gap between the degenerate ground 
states and the excited state, and $d_\ell^+$, $d_\ell^-$ be the dipole moments of the 
$\sigma_+$ and $\sigma_-$ transition, respectively, for the $\ell$th dot. Without loss
of generality we can assume $\op{H}_m=\oplus_{\ell=1}^L \op{H}_{m,\ell}$ for $m=0,1,2$, 
$\ell=1,\ldots,L$, and
\[ \fl
  \op{H}_{0,\ell} = \frac{\eps_\ell}{3} 
  \left( \begin{array}{ccc} -1 & 0 & 0 \\ 0 & -1 & 0 \\ 0 & 0 & +2 \end{array}\right), 
  \;
  \op{H}_{1,\ell} = d_\ell^+ 
  \left( \begin{array}{ccc} 0 & 0 & 1 \\ 0 & 0 & 0 \\ 1 & 0 & 0 \end{array}\right), 
  \;
  \op{H}_{2,\ell} = d_\ell^- 
  \left( \begin{array}{ccc} 0 & 0 & 0 \\ 0 & 0 & 1 \\ 0 & 1 & 0 \end{array}\right).
\]
The ensemble is simultaneously (mixed-state) controllable exactly if the Lie algebra
generated by $\{\op{H}_m\}$ is $\oplus_{\ell=1}^L \su(3)$.  Again, this is usually the
case unless two or more dots in the ensemble are effectively identical.  For example, 
we verified that even if $d_\ell^+=1$ and $d_\ell^-=-1$ for all $\ell$, the Lie algebra
for a 5-dot ensemble with $\eps_\ell=1+\Delta\eps_\ell\ge 0$ and $\Delta\eps_\ell \neq 
\Delta\eps_{\ell'}$ unless $\ell=\ell'$ had indeed dimension $40 = 5 \times (3^2-1)$. 
However, if two dots have the same energy gap $\eps_\ell$ and the same (absolute) values
of the dipole moments $d_\ell^\pm$ then simultaneous controllability is lost, e.g., if 
we have $\eps_1=\eps_4$ and $d_4^\pm=d_1^\pm$ or $d_4^\pm=-d_1^\pm$ then the Lie algebra
dimension is only $32=4 \times (3^2-1)$.  However, if $\eps_1=\eps_4$ but $d_4^+\neq 
\pm d_1^+$, for instance, then simultaneous controllability is maintained.  

In the previous example, we can also ask what about controllability if we can only 
apply a single pulse consisting of a mixture of $\sigma_+$ and $\sigma_-$ polarized 
light.  In this case we replace $\op{H}_1$ and $\op{H}_2$ by the composite control 
Hamiltonian $\op{H}_C=\cos\alpha\op{H}_1+\sin\alpha\op{H}_2$ with $\alpha\in [0,\pi/2]$,
so that the single dot Hamiltonians are
\[ \fl
  \op{H}_{0,\ell} = \frac{\eps_\ell}{3} 
  \left( \begin{array}{ccc} -1 & 0 & 0 \\ 0 & -1 & 0 \\ 0 & 0 & +2 \end{array}\right), 
  \quad
  \op{H}_{C,\ell} = 
  \left( \begin{array}{ccc} 0 & 0 & d_\ell^+\cos\alpha \\ 
                            0 & 0 & d_\ell^-\sin\alpha \\ 
                            d_\ell^+\cos\alpha & d_\ell^-\sin\alpha & 0 
         \end{array}\right).
\]
We see clearly that the system cannot be simultaneously controllable for $\alpha=0$ or
$\alpha=\pi/2$ as in this case none of dots are individually controllable, their Lie 
algebras being contained in $\u(2)$.  However, if we assume $d_\ell^-=-d_\ell^+$, which
is often the case in practice, then even for $\alpha \in (0,\pi/4)$, the Lie algebra 
generated by $\op{H}_{0,\ell}$ and $\op{H}_{C,\ell}$ above is a unitary representation 
of $\so(3)\oplus \u(1)$.  Hence, the dots are not individually controllable, and the
ensemble is thus not simultaneously controllable.  The Lie algebra of an ensemble of 
$L$ (non-identical) dots of this form is a unitary representation of $(\oplus_{\ell=1}^L 
\so(3)) \oplus \u(1)$.  Note that $\so(3)$ is \emph{not} even sufficient for pure-state
controllability.  If $d_\ell^-\neq \pm d_\ell^+$ then the $\ell$th dot is generally 
invidually controllable for $\alpha \in (0,\pi/2)$ and any ensemble of non-identical 
dots would be simulataneously controllable with a mixed-polarization control pulse.

\section{Optimal Control of Ensembles of Quantum Dots}
\label{sec:dots_optimal}

Simultaneous controllability of an ensemble of non-interacting quantum dots implies 
in particular that it is possible to selectively excite a particular dot with a single
laser pulse without the need for selective addressing.  If the energy levels of the
dots, and hence their resonance frequencies, are different, a standard approach would
be to use frequency-selective addressing using simple, e.g., Gaussian pulses resonant 
with the transition frequency of the dot to be excited.  However, this may be less 
than optimal, especially when the pulse length is to be kept to a minimum to achieve 
fast operations, which would be crucial in quantum information processing applications. 
The question therefore naturally arises whether the results could be improved using 
optimally shaped pulses.

To address this question, we consider an ensemble of five quantum dots, modelled as 
two-level systems with energy differences $\eps_\ell$ of $1.32$, $1.35$, $1.375$, 
$1.38$ and $1.397$~eV, respectively.  If interdot coupling is negligible, the internal 
system Hamiltonian is
\begin{equation}
  \op{H}_0       = \diag(\op{H}_{0,1},\ldots,\op{H}_{0,5}), \quad
  \op{H}_{0,\ell} = \eps_\ell \sz/2, 
\end{equation}
and the Hamiltonian that describes the coupling to the external driving field is 
\begin{equation}
 \op{H}_1       = \diag(\op{H}_{1,1},\ldots,\op{H}_{1,5}), \quad
 \op{H}_{1,\ell} = d_\ell \sx 
\end{equation}
where $d_\ell$ is the dipole coupling of the $\ell$th dot to the field.  Even if we
assume, for simplicity, that all the dipole couplings are equal $d_\ell=1$ for $\ell=1,
\ldots,5$, the dots are still simultaneously controllable, and in particular we can 
selectively excite a single dot without spatial addressing.  

\begin{figure}
(a) \myincludegraphics[width=4.5in]{figures/pdf/QD1a-trajct.pdf}{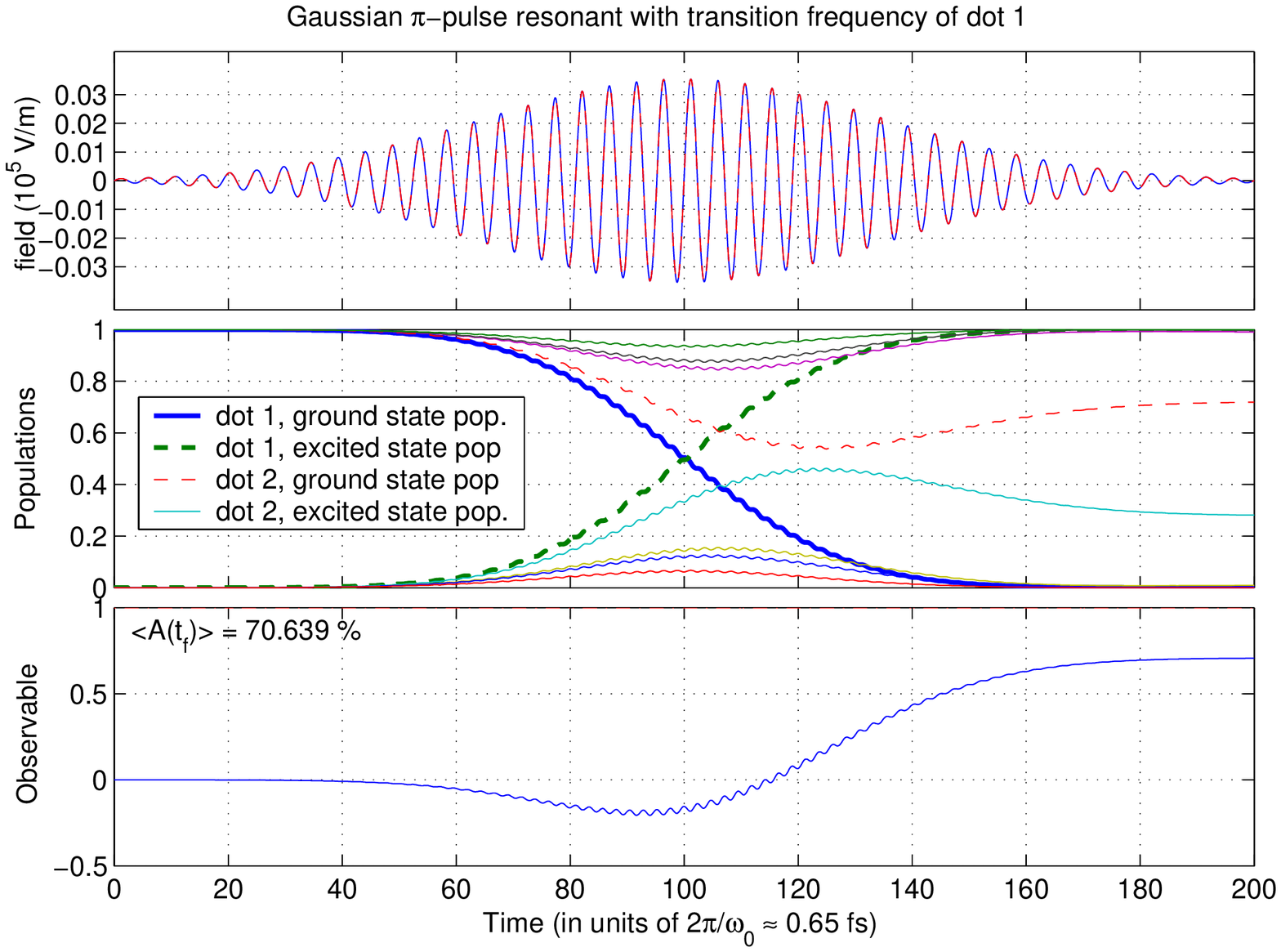} \\
(b) \myincludegraphics[width=4.5in]{figures/pdf/QD1-trajct.pdf}{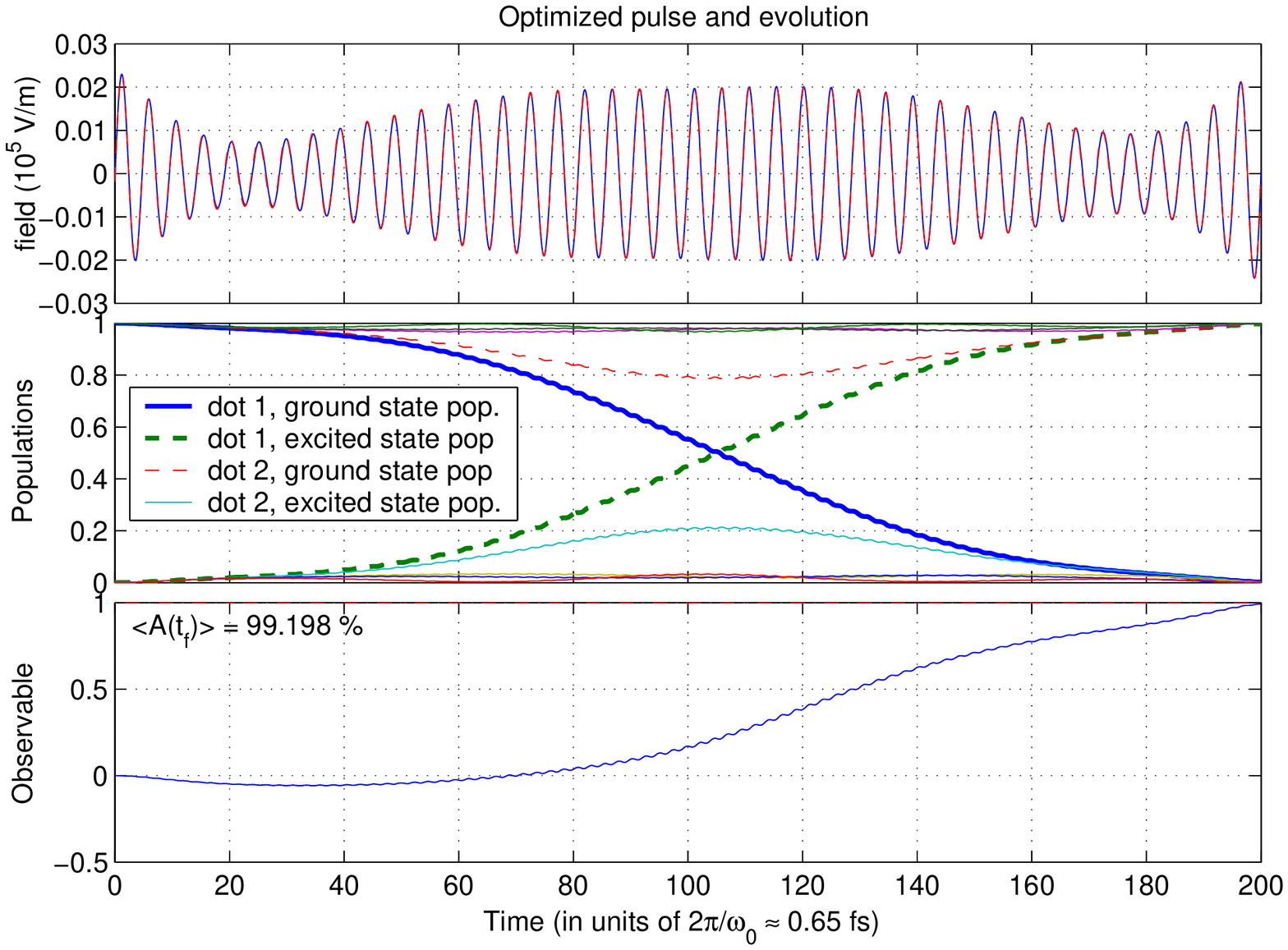} 
\caption{Selective excitation of dot 1: control field (in units of $10^5$~V/m) and 
evolution of the populations and observable for (a) a frequency-selective Gaussian 
control pulse, and (b) an optimally shaped control pulse.  The expectation value of 
the observable corresponds to the population of the excited state of the target dot 
\emph{minus} the sum of the populations of the excited states of \emph{all} other dots.  
The Gaussian pulses (a) 
accomplishes the goal of exciting the first dot but also leads to significant excitation
of the other dots, especially the dot energetically closest to the target dot, at the 
target time, while the shaped pulse (b) ensures that the excited state populations of 
all dots except the target dot return to (almost) zero at the target time.}
\label{fig3}
\end{figure}

Fig.~\ref{fig3}(a) shows the evolution of the ground and excited state populations of
the dots if we simply apply a $\pi$-pulse, resonant with the transition frequency of 
the first dot, with Gaussian envelope $A(t)= q\sqrt{\pi}\exp[-q^2(t-t_f/2)^2]$, where 
$q=4/t_f$ and the target time is $t_f=200$ time units ($\approx 130$~fs). Although the
pulse achives almost 100\% population transfer from the ground to the excited state 
for the target dot, it also leads to significant unwanted excitation of energetically
adjacent dots.  This effect tends to become more pronounced the shorter the pulses, 
and the smaller the differences in the transition frequencies of the dots.

Fig.~\ref{fig3}(b) shows that we can considerably improve the results in this case 
using shaped pulses.  The shaped pulse still achieves near perfect excitation of the
target dot but considerably reduces the overall excitation of the other dots.  Most
importantly though, while there is some remaining transient excitation of the other 
dots, the shaped pulse ensures that the populations of the excited states return to 
(almost) zero at the target time, except for the target dot, for which the excited
state population is almost~1.

The pulse shown in Fig.~\ref{fig3}(b) was obtained using an iterative optimal control
algorithm similar to~\cite{PRA61n012101}.  The starting point for the algorithm was 
the Gaussian pulse shown in Fig.~\ref{fig3}.  The observable to be optimized was chosen
to be $\op{A}=\diag(\op{P},\op{Q},\op{Q},\op{Q},\op{Q})$, where $\op{P}=\ket{1}\bra{1}$ 
is the projection onto the upper level and $\op{Q}=-\op{P}$ to reflect our objective to
at once maximize the excited state population of the first dot, while minimizing the 
excitation of all other dots.  Note that this is not the only possible choice of the 
observable but the results of the algorithm depend significantly on the choice of the 
target observable.  For example, choosing $\op{A}'=\diag(\op{P},\op{0},\op{0},\op{0},
\op{0})$ would be a bad choice and unlikely to improve the results because the initial
Gaussian pulse does accomplish the objective of achieving 100\% population transfer for
the target dot, and the populations of the other dots do not affect the expectation 
value of $\op{A}'$, whence there would be no reason for the algorithm to alter the
pulse shape in an attempt to suppress the off-resonant excitation of the other dots.

\section{Conclusion}

We have considered various notions of controllability for quantum systems and their
application to multi-partite systems.  In particular, we have investigated the degree
of simultaneous controllability of the components of a decomposable system by means 
of global control, and given \emph{necessary} and \emph{sufficient} criteria for the 
simultaneous mixed-state and pure-state controllability of the components.  The results
have been applied to the problem of simultaneous controllability of quantum dots with
negligible interdot coupling, and in particular the problem of selective excitation of
ensembles of quantum dots with a laser pulse that is applied to the entire ensemble.  
We have also shown, for a simple model, that shaped pulses derived using optimal control
theory may offer substantial improvements in selectivity compared to Gaussian pulses.

\ack
SGS would like to thank Andrey Bychkov for helpful discussions and suggestions and
acknowledges financial support from Fujitsu, the Cambridge-MIT Institute Quantum
Technology Project, and the EPSRC.

\appendix
\section{Proof of theorem \ref{thm:simCtrl}}
\label{app:proof}

Assume $\op{H}[\vec{f}(t)]$ of the form~(\ref{eq:Ht}), and $\op{\rho}(t)$ and $\op{H}_m$ 
of the form~(\ref{eq:Hm}).  For $m=0,\ldots,M$ and $\ell=1,\ldots,L$, set $\alpha_{m,\ell}
=\Tr(\op{H}_{m,\ell})$ and define the diagonal generators
\begin{equation} \label{eq:Dm}
   \tilde{D}_m = \diag(\alpha_{m,1}\op{I}_{N_1},\ldots,\alpha_{m,L} \tilde{I}_{N_L}).
\end{equation}
where $\op{I}_\ell$ is the identity matrix in dimension $\ell$, and the trace-zero
generators
\begin{equation} \label{eq:tHm}
  \tilde{H}_m =  \diag(\tilde{H}_{m,1},\ldots,\tilde{H}_{m,L}),
\end{equation}
where the blocks on the diagonal are given by
\begin{equation} \label{eq:tHml}
  \tilde{H}_{m,\ell} = \op{H}_{m,\ell}-\frac{\alpha_{m,\ell}}{N_\ell} \op{I}_\ell.
\end{equation}
The diagonal elements $\op{D}_m$ commute with $\op{\rho}(t)$ for all $m=0,\ldots,M$.
Hence, the orbits of $\op{\rho}(t)$ generated by $\op{H}[\vec{f}(t)]$ and $\tilde{H}
[\vec{f}(t)]=\tilde{H}_0+\sum_{m=1}^M f_m(t) \tilde{H}_m$ are identical.  

Let $\tilde{\L}=\L(\{\tilde{H}_m\})$ be the Lie algebra generated by the trace-zero
skew-Hermitian matrices $\rmi\tilde{H}_m$.  Due to the structure of the generators, 
the Lie algebra $\tilde{\L}$ must be a subalgebra of $\oplus_{\ell=1}^L \su(N_\ell)$,
and it follows from classical results~\cite{72JS} that the orbits of a generic mixed
state~\cite{JPA37p1389} will be maximal if and only if $\tilde{\L}=\oplus_{\ell=1}^L 
\su(N_\ell)$.

Now let $\L=\L(\{\op{H}_m\})$ be the Lie algebra generated by $\rmi\op{H}_m$, and $\L_D
=\L(\{\op{D}_m\})$ be the Lie algebra generated by the $M+1$ diagonal (skew-Hermitian) 
matrices $\rmi\op{D}_m$.  Since the diagonal generators $\rmi\op{D}_m$ commute with the 
trace-zero matrices $\tilde{H}_m$, i.e., $[\op{D}_m,\tilde{H}_{m'}]=0$ for all $m,m'=
1,2,\ldots,M$, we have
\begin{equation}
   \L = \tilde{\L} \oplus \L_D
\end{equation}
i.e., the Lie algebra $\L$ is the direct sum of the Lie algebras $\tilde{\L}$ and 
$\L_D$.  Noting that the dimension of $\su(N_\ell)$ is $N_\ell^2-1$ and the dimension 
of $\L_D$ is equal to the rank of the matrix
\begin{equation} \label{eq:A}
  A = 
  \left( \begin{array}{ccc}  
  \alpha_{0,1} & \ldots & \alpha_{0,L} \\
  \vdots       &        & \vdots \\
  \alpha_{M,1} & \ldots & \alpha_{M,L} 
  \end{array} \right)
\end{equation}
where $r=\mbox{rank}(A) \le \min(M+1,L)$, we obtain the Lie algebra dimension condition
of Theorem~\ref{thm:simCtrl} as a necessary and sufficient condition for simultaneous 
(mixed-state) controllability.

\section{Controllability analysis for two uncoupled two-level systems}
\label{app2}

To make the difference between individual and simultaneous controllability explicit,
consider two non-interacting two-level systems simultaneously driven by a coherent
control field $f(t)$.  We can write the Hamiltonian of the composite system as 
$\op{H} = \op{H}_0+f(t)\op{H}_1$, where
\begin{eqnarray*}
  \op{H}_0   &=& \diag(E_0^{(1)},E_1^{(1)}) \oplus \diag(E_0^{(2)},E_1^{(2)}) \\
  \op{H}_1   &=& d_1 \sx \oplus d_2 \sx.
\end{eqnarray*}
$\op{H}_0$ can further be split into a trace-zero part $\op{H}_0'=\eps_1\sz \oplus
\eps_2 \sz$, where $\eps_\ell=(E_1^{(\ell)}-E_0^{(\ell)})/2$, and a diagonal part
$\op{D}_0=\alpha_1\op{I}\oplus\alpha_2 \op{I}$, where $\alpha_\ell=(E_0^{(\ell)}+
E_1^{(\ell)})/2$.  If $\eps_\ell=0$ or $d_\ell=0$ then subsystem $\ell$ is not even
individually controllable.  Hence, we shall assume that $\eps_\ell \neq 0$ and $d_\ell
\neq 0$ for $\ell=1,2$. 

The diagonal generator $\op{D}_0$ is not relevant for our controllability analysis.
The traceless generators $W_1=\rmi\op{H}_0'/\eps_1$, $W_2=\rmi\op{H}_1/d_1$ give 
rise to the following Lie algebra:
\begin{eqnarray*}
  W_3 &= [W_2,W_1]/2 &= -i(\sy \oplus a   b   \sy) \\
  W_4 &= [W_3,W_1]/2 &= -i(\sx \oplus a^2 b   \sx) \\
  W_5 &= [W_3,W_2]/2 &=  i(\sz \oplus a   b^2 \sz) \\
  W_6 &= [W_4,W_1]/2 &=  i(\sy \oplus a^3 b   \sy) \\
  W_7 &= [W_3,W_4]/2 &= -i(\sz \oplus a^3 b^2 \sz) 
\end{eqnarray*}
where $a = \eps_2/\eps_1$ and $b= d_2/d_1$.  Combining these terms yields
\begin{eqnarray*}
  W_2+W_4     &= 0 \oplus (1-a^2)b \, i\sx \\
  a^2 W_2+W_4 &= (a^2-1) \, i\sx \oplus 0 \\
  W_3+W_6     &= 0 \oplus ab(a^2-1)\, i\sy \\ 
  a^2 W_3+W_6 &= (1-a^2) \, i\sy \oplus 0 \\
  W_5+W_7     &= 0 \oplus ab^2 (1-a^2) \, i\sz \\
  a^2 W_5+W_7 &= (a^2-1) \, i\sz \oplus 0   
\end{eqnarray*}
This shows that if $a,b\neq 0,\pm 1$ then the relevant Lie algebra $\tilde{\L}$ of the
system is $\su(2)\oplus \su(2)$.  

For $\alpha=\pm 1$ we can similarly show that $V_k=W_k$ for $k=1,2,3$, $V_4=W_4$ and
\begin{eqnarray*}
  V_5 &= [V_2,V_4]/2 &= \sy \oplus ab^3 \sy \\
  V_6 &= [V_4,V_3]/2 &= i(\sx \oplus b^3 \sx) \\
  V_7 &= [V_2,V_5]/2 &= i(\sz + ab^4 \sz).
\end{eqnarray*}
Combining these terms leads to
\begin{eqnarray*}
  V_6-V_2     &= 0 \oplus b(b^2-1) i\sx \\
  b^2 V_2-V_6 &= (b^2-1) i\sx \oplus 0 \\
  V_5+V_3     &= 0 \oplus ab(b^2-1) \sy \\ 
  b^2 V_3-V_5 &= (b^2-1) \sy \oplus 0 \\
  V_7-V_4     &= 0 \oplus ab^2 (b^2-1) i\sz \\
  b^2 V_4-V_7 &= (b^2-1) i\sz \oplus 0   
\end{eqnarray*}
This shows that if $a\neq 0$, $b\neq 0,\pm 1$ then the relevant Lie algebra $\tilde{\L}$ 
of the system is still $\su(2)\oplus \su(2)$, and the two subsystems are therefore 
simultaneously controllable.

However, if $a=\pm 1$ and $b=\pm 1$ then $V_4=2V_1$, $V_5=V_3$, $V_6=V_2$, $V_7=V_1$,
i.e., Lie algebra is three-dimensional, and since $V_3=[V_2,V_1]$, a representation of
$\su(2)$.  The non-interacting two-level systems are therefore individually but not
simultaneously controllable.

\section*{References}

\bibliography{papers,papers_old,sonia,books}
\bibliographystyle{prsty}
\end{document}